\documentclass[doublespacing]{elsart}
\usepackage{icarus}

\usepackage{natbib}

\bibpunct{(}{)}{;}{a}{,}{,}

\usepackage{graphicx}
\usepackage{lscape}
\usepackage{longtable}
\usepackage{CJKutf8}

\newcommand{\BibTeX}{ \textrm{B\kern-.05em\textsc{i\kern-.025em b}\kern-.08em
    T\kern-.1667em\lower.7ex\hbox{E}\kern-.125emX} }

\newcounter{ionctr}
\ifx \undefined \ion \newcommand{\ion}[2]{\setcounter{ionctr}{#2}{#1$\;${\small\rmfamily\Roman{ionctr}}\relax}} \fi

\begin{document}

\begin{frontmatter}

% Title, authors and addresses

% use the thanksref command within \title, \author or \address for footnotes;
% use the corauthref command within \author for corresponding author footnotes;
% use the ead command for the email address,
% and the form \ead[url] for the home page:
% \title{Title\thanksref{label1}}
% \thanks[label1]{}
% \author{Name\corauthref{cor1}\thanksref{label2}}
% \ead{email address}
% \ead[url]{home page}
% \thanks[label2]{}
% \corauth[cor1]{}
% \address{Address\thanksref{label3}}
% \thanks[label3]{}

\title{Observational Constraints on Water Sublimation from 24 Themis and 1 Ceres}
% use optional labels to link authors explicitly to addresses:
% \author[label1,label2]{}
% \address[label1]{}
% \address[label2]{}

\author[label1]{Adam J. McKay},
\author[label2]{Dennis Bodewits},
\author[label2]{Jian-Yang Li}

\address[label1]{Univerisity of Texas Austin/McDonald Observatory, 2512 Speedway Stop C1402, Austin, TX 78712, (U.S.A);amckay@astro.as.utexas.edu}
\address[label2]{Department of Astronomy, University of Maryland, College Park, MD 20742-2421 (U.S.A.); dennis@astro.umd.edu, jyli@astro.umd.edu}

%% This copyright statement isn't required at any stage by the Icarus
%% Editorial Office or Elsevier.  However, until you sign over the
%% copyright to Elsevier prior to publication (or negotiate with them
%% about copyright), you own the copyright to anything you create.
%% Just to keep things unambiguous, always include a copyright statement
%% or explicitly dedicate your work to the public domain.
\begin{center}
\scriptsize
Copyright \copyright\ 2016 Adam J. McKay,  Dennis Bodewits, Jian-Yang Li
\end{center}

%% ----- ELSEVIER STUFF -----
%% The commands below up to the \end{frontmatter} are commented out
%% so that we can do some Icarus-required formatting on the second and
%% third pages that is not required later on by Elsevier.  So when
%% your paper gets accepted, and you are ready to start dealing with
%% Elsevier, copy your abstract and keywords up here, uncomment these
%% lines, and comment out the ICARUS STUFF below.
%% 
%% Alternately, you might just want to move these abstract, keyword,
%% and end frontmatter commands down, and comment out the ICARUS STUFF
%% commands.  It doesn't matter.

% \begin{abstract}
% % Text of abstract
% 
% \end{abstract}
% 
% \begin{keyword}
% % keywords here, in the form: keyword \sep keyword
% 
% 
% % PACS codes here, in the form: \PACS code \sep code
% 
% \end{keyword}

%% ----- END ELSEVIER STUFF -----

\end{frontmatter}

%% ----- ICARUS STUFF -----
%% Some formatting on the first, second, and third pages are required
%% by the Icarus Editorial Office that are not required by Elsevier.
%% This section contains those things.  When you are ready to transition
%% to ``Elsevier'' mode, copy your abstract and keywords up into
%% the ELSEVIER STUFF section, and then you can just delete everything
%% in this section.

%% We need to list the number of manuscript pages, figures, and tables. 
%%
%% Rather than manually count these things out, we'll use a little
%% trick here from Paul.  All you have to do is place three \label{}
%% tags on the last page, the last table, and the last figure, that
%% way these values are automatically updated (as long as you remember
%% to move the lasttable and lastfig labels when you add or remove
%% tables and figures).

\begin{flushleft}
\vspace{1cm}
Number of pages: \pageref{lastpage} \\
Number of tables: \ref{lasttable}\\
Number of figures: \ref{lastfig}\\
\end{flushleft}

%% Don't worry about finding the various last* tags and deleting them
%% when you go to ``Elsevier'' mode if you don't want to, they should be
%% silently ignored.

%% The second page should indicate a proposed running head of not more 
%% than 55 characters, and the name and address to which editorial 
%% correspondence and proofs should be directed.  The pagetwo 
%% environment that icarus.sty provides will make page two for you,
%% just give the running head as an argument to the environment, and
%% then your correspondence address inside.
\begin{pagetwo}{Water Sublimation from Themis and Ceres}

Adam J. McKay \\
University of Texas Austin\\
2512 Speedway, Stop C1402\\
Austin, TX 78712, USA. \\
\\
Email: amckay@astro.as.utexas.edu\\
Phone: (512) 471-6493

\end{pagetwo}

%\begin{CJK}{UTF8}{bsmi}
%文章内容
%\end{CJK}

\begin{abstract}
Recent observations have suggested that there is water ice present on the surfaces of 24 Themis and 1 Ceres. We present upper limits on the H$_2$O production rate on these bodies derived using a search for [\ion{O}{1}]6300~\AA~emission.  For Themis, the water production is less than 4.5 $\times$ 10$^{27}$ mol s$^{-1}$, while for Ceres our derived upper limit is 4.6 $\times$ 10$^{28}$ mol s$^{-1}$. The derived limits imply a very low fraction of the surface area of each asteroid is active ($<$ 2 $\times$ 10$^{-4}$), though this estimate varies by as much as an order of magnitude depending on thermal properties of the surface. This is much lower than seen for comets, which have active areas of 10$^{-2}$ - 10$^{-1}$. We discuss possible implications for our findings on the nature of water ice on Themis and Ceres.
\end{abstract}

% %% Keywords should appear after the abstract. 

%% ----- END ICARUS STUFF -----

%main text
\section{Introduction}
\indent Recent observations have provided evidence that several asteroids in the Main Belt contain water ice. The discovery of Main Belt Comets (MBCs, also termed Active Asteroids) has shown that some asteroids in the Main Belt exhibit cometary activity in the form of dust comae and tails. While the apparent activity for some MBCs (e.g. 596 Scheila, 311P/PanSTARRS) has been shown to be caused by collisions or other non-sublimation processes~\citep{Bodewits2011, Jewitt2013}, the recurrent activity observed for objects such as 133P/Elst-Pizarro, 238P/Read, and 313P/Gibbs, suggests that this activity is driven by sublimation of water ice~\citep{Hsieh2010, Hsieh2011, Hsieh2015, Jewitt2015}, though direct detection of any gas around MBCs has proven
elusive~\citep[e.g.][]{Licandro2011, deValBorro2012, ORourke2013}.\\

\indent However, water ice has been reportedly detected on the surfaces of 24 Themis and 65 Cybele~\citep{Campins2010, RivkinEmery2010}, as well as in Oxo Crater on 1 Ceres~\citep{Combe2016}.  H$_2$O gas was recently detected around Ceres~\citep{Kueppers2014}, as well as a more tentative detection several decades ago~\citep{AHearnFeldman1992}.  Several MBCs are in the Themis dynamical family, marking a possible link between the surface ice detected on Themis and the activity observed around these MBCs. Despite the evidence for water ice in the Main Belt, the exact properties of this ice remain poorly understood. Models suggest that it must be buried beneath the surface regolith to be stable over the age of the Solar System~\citep[e.g.][]{FanaleSalvail1989, Capria2012} though at the same time observations of surface ice on Themis and Cybele suggest that it does make its way to the surface.  Recent results from the gRaND instrument on Dawn show a higher hydrogen abundance at higher latitudes, suggesting there is more buried ice at higher latitudes~\citep{Prettyman2016}.  To date, water vapor has only been detected around Ceres. The detection of water outgassing from additional asteroids with suspected water ice on their surfaces would both confirm the presence of water ice on these bodies as well as provide constraints on the nature and
distribution of that ice (surface vs. subsurface, uniform distribution or isolated patches, pure ice vs. dirty ice mixture, etc.).\\

\indent A sensitive probe for H$_2$O in the gas phase at optical wavelengths is the forbidden oxygen line at 6300.3~\AA. This line results from prompt emission of atomic oxygen after a photodissociation event. This line has been employed in
observations of comets as a proxy of the H$_2$O production rate~\citep[e.g.][]{FinkHicks1996, Morgenthaler2001, Morgenthaler2007, McKay2014, McKay2015}. We present a search for [\ion{O}{1}]6300 emission around 24 Themis and 1 Ceres in an
effort to constrain the outgassing of H$_2$O from their surface/subsurface. Section 2 describes the observations and our reduction and analysis procedures.  Section 3 presents our derived upper limits on the H$_2$O production rate, and
section 4 discusses the implications of these results for the nature of the ice present on Themis and Ceres. We summarize our findings in Section 5.

\section{Observations and Data Analysis}
\subsection{Observations}
\indent We obtained spectra with the ARCES echelle spectrometer, mounted on the Astrophysical Research Consortium 3.5-m telescope at Apache Point Observatory (APO) in Sunspot, New Mexico. ARCES provides a spectral resolution of R $\equiv$
$\frac{\lambda}{\Delta\lambda}$ = 31,500 and a spectral range of 3500-10,000~\AA~with no interorder gaps. More specifics for this instrument are discussed elsewhere~\citep{Wang2003}. The observation dates and geometries are described in Table 1.  Both asteroids were observed less than a year before their perihelion.  We used an ephemeris generated from JPL Horizons for non-sidereal tracking.  For short time-scale tracking, the guiding software uses a boresight technique, which utilizes optocenter flux that falls outside the slit to keep the slit on the optocenter. We observed a G2V star in order to remove the underlying solar continuum and Fraunhofer absorption lines. We obtained spectra of a
fast rotating (vsin(i) $>$ 150 km s$^{-1}$), O, B, or A star to account for telluric features and spectra of a flux standard to convert observed counts to absolute flux. The calibration stars used for each observation date are given in Table~\ref{observations}.  We obtained spectra of a quartz lamp for flat fielding and acquired spectra of a ThAr lamp for wavelength calibration.

\subsection{Data Reduction and Analysis}

\indent Spectra were extracted and calibrated using IRAF scripts that perform bias subtraction, cosmic ray removal, flat fielding, and wavelength calibration. We removed telluric absorption features, the reflected solar continuum from the
surface, and flux calibrated the spectra employing our standard star observations. We assumed an exponential extinction law and extinction coefficients for APO when flux calibrating the asteroid spectra~\citep{Hogg2001}. More details of this procedure can be found in~\cite{McKay2012}. We determined slit losses by performing aperture photometry on the slit viewer images as
described in~\cite{McKay2014}.\\

\indent We determined H$_2$O production rates from our [\ion{O}{1}]6300~\AA~line fluxes by employing a Haser model modified to emulate the more physical vectorial model. More details of the model can be found in~\cite{Morgenthaler2001, Morgenthaler2007} and~\cite{McKay2014, McKay2015}.

\section{Results}
\indent We display the spectral region containing the [\ion{O}{1}]6300~\AA~line for our Themis and Ceres spectra in Fig.~\ref{spectra}. Neither object shows evidence for [\ion{O}{1}]6300~\AA~emission. We therefore calculate 3-sigma upper limits on the [\ion{O}{1}]6300~\AA~flux and the corresponding upper limit on the H$_2$O production rate, which are shown in Table~\ref{upperlimit}. These upper limits reflect the line flux for a Gaussian line profile of the instrumental line width with a peak intensity that is 3 times the local noise.  For Ceres, there is clearly systematic scatter due to the subtraction of a very strong continuum, so in this case we adopted the standard deviation of the continuum subtracted spectra in the neighborhood of the expected [\ion{O}{1}]6300~\AA~emission as the local 1-sigma noise.  We derive a 3-sigma upper limit on the H$_2$O production rate of 4.5 $\times$ 10$^{27}$ mol s$^{-1}$ for Themis and 4.6 $\times$ 10$^{28}$ mol s$^{-1}$ for Ceres. Our upper limit is much more sensitive for Themis than for Ceres because Themis is a much fainter object (V=12 at the time of observation) than Ceres (V=8), meaning the Ceres spectra contain a much higher continuum level that is more effective at concealing any [\ion{O}{1}]6300~\AA~emission that may be present (in addition to the systematic scatter discussed above).\\

\indent Using the sublimation model of~\cite{CowanAhearn1979}, we can convert the upper limit on H$_2$O production to an upper limit on the fraction of each asteroid’s surface that is actively sublimating. We present the upper limit on the active fraction for different thermal models in Table~\ref{activefraction}. The isothermal model assumes that the whole surface is the same temperature while the subsolar model assumes the whole surface has a temperature equal to the subsolar
point. The isothermal and subsolar models are not realistic, but do provide upper (isothermal) and lower (subsolar) bounds to the limits we can place on the active fraction. The fast-rotator model assumes that the lines of latitude are isotherms, which is expected if the temperature at each latitude is determined by the average diurnal insolation at that latitude. This is expected to be the case if the rotation rate is fast or the thermal inertia of the surface material is high. The slow-rotator model assumes that the temperature of each surface element is in equilibrium with the instantaneous solar flux incident
at that location. This approximation has been used to estimate the active areas of cometary nuclei~\citep[e.g.][]{Ahearn1995,Bodewits2014}. As neither Themis nor Ceres has a slow rotation period (P $<$ 10 hours), this assumption is appropriate only if the thermal inertia of the surface is very low.  The thermal inertia of Ceres has been measured to be less than 15 tiu~\citep{Chamberlain2009}.  No constraints on the thermal inertia of Themis are available.  However, large asteroids such as Themis tend to have low thermal inertias~\citep{Delbo2015}.  Therefore we believe that the slow-rotator approximation is the most realistic approximation for these asteroids.\\

\indent For the heliocentric distances of Ceres and Themis at the time of observation and adopting bond albedos of 0.02 and 0.03 for Themis and Ceres, respectively, we calculate the sublimation rate of H$_2$O per unit surface area for each model, which are given in Table~\ref{activefraction}.  We calculated the bond albedo by the following:
\begin{equation}
A_{bond}=qA_{geo} 
\end{equation}
where $A_{geo}$ is the geometric albedo, taken as 0.07 for Themis~\citep{Masiero2011} and 0.09 for Ceres~\citep{Li2006}, and $q$ is the phase integral.  We assume a phase integral of 0.3 for Themis and Ceres, which is typical of C-type asteroids~\citep{Li2015}.
By dividing our derived upper limit on the H$_2$O production by the expected sublimation rate, we derive the maximum active area of the surface. Then we divide this active area by the total surface area of each asteroid to get an upper limit on the fraction of the surface area that is active.\\

\section{Discussion}
\indent Since both Themis and Ceres are much more massive than a typical comet, it is possible that a large fraction of any gas present could be gravitationally bound to the asteroid, meaning the Haser-derived gas production rate is not indicative of the sublimation occuring on the surface. However, for Themis's escape velocity of 90 m s$^{-1}$ and a gas temperature of 160 K (i.e. the expected surface temperature of Themis), less than 1\% of the gas molecules have velocities less than the escape velocity (assuming a Maxwellian velocity distribution), meaning that the assumption of radial outflow is a reasonable approximation. For Ceres, with an escape velocity of 500 m s$^{-1}$, 66\% of the particles have velocities less than the escape velocity.  This means that any water vapor present would be more concentrated around the asteroid than in the Haser model. This means that our upper limit should actually be lower than the Haser model would indicate. However, as \ion{O}{1} is a photodissociation product, the excess velocity imparted on the atom after photodissociation~\citep[$\sim$ 1 km s$^{-1}$,][]{WuChen1993}, means that a majority of the \ion{O}{1} will achieve escape velocity. Additionally, as previous works have assumed that the Haser model is valid~\citep{JewittGuilbert2012, Kueppers2014}, for comparison to previous work the Haser model derived production rates are the most appropriate.\\

\indent For Ceres, H$_2$O vapor was detected by the Herschel Space Observatory with a production rate of approximately 2 $\times$ 10$^{26}$ mol s$^{-1}$~\citep{Kueppers2014}. Our upper limit is not extremely sensitive and is fully consistent with the Herschel derived value.  The detections reported in~\cite{Kueppers2014} occurred in October 2012 and March 2013, bracketing our observations.  Our observations cover a range in sub-observer latitude on Ceres of approximately 100-310$^{\circ}$, which encompasses the most active longitudes of 120$^{\circ}$ and 240$^{\circ}$ reported by~\cite{Kueppers2014}.~~\cite{Kueppers2014} derive an active fraction of approximately 10$^{-7}$, consistent with our upper limits on the active fraction regardless of the thermal model used.  This limit is also consistent with the current ice coverage on Ceres as observed by Dawn~\citep{Combe2016}.\\

\indent For Themis, water vapor has never been detected, but several upper limits have been derived.~\cite{JewittGuilbert2012} employed optical spectra to search for emission due to CN, a molecule that is commonly observed in traditional comets. Assuming the average CN/H$_2$O ratio of 0.3\% observed in comets, they derive an upper limit of 1.3 $\times$ 10$^{28}$ mol s$^{-1}$, though this upper limit should be treated with some caution as it assumes that any water ice present on Themis has a cometary CN/H$_2$O ratio. Using observations of OH, which is released from H$_2$O photodissociation and thus can be used as a direct tracer of H$_2$O,~\cite{Lovell2010} derived an upper limit of $\sim$ 10$^{28}$ mol s$^{-1}$.  Our upper limit of 4.5 $\times$ 10$^{27}$ mol s$^{-1}$ provides the most stringent constraint on H$_2$O production obtained for Themis so far.  A potentially more sensitive upper limit may become available from Herschel observations (O'Rourke et al. in prep. and private communication).\\

\indent Our upper limit on the H$_2$O production rate (and those of previous works as well) suggests that a very low fraction of Themis's surface is actively sublimating. As water ice was detected at all epochs over several years of observations, both~\cite{Campins2010} and~\cite{RivkinEmery2010} suggest that the water ice layer coats the entire surface, which is being replenished from the subsurface. Therefore this uniform coating should be actively sublimating, yielding an active fraction near unity. This is inconsistent with our upper limit, which implies a very small active fraction. This conclusion is independent of the thermal model adopted, as no thermal model provides an active fraction near unity (see Table~\ref{activefraction}).\\

\indent This discrepancy could be due to several factors. One possibility is that the ice is buried under an insolating layer of regolith, but this is unlikely as infrared observations only probe microns deep into the surface, which is not likely deep enough to keep any water ice present from sublimating. As activity could be intermittent, our observations could have occurred at a time when any water ice is not actively sublimating, but this would suggest a non-uniform distrubution of water ice on the surface, inconsistent with the near uniform distribution implied by IR observations of surface ice. It has also been claimed
that the spectral signature observed by~\cite{Campins2010} and~\cite{RivkinEmery2010} is not due to water ice but to the mineral goethite, meaning there is no water ice to sublimate~\citep{Beck2011}. However, it has been argued by other authors that this is unlikely as goethite in meteorites is a result of aqueous alteration after fall and this mineral has not been detected in freshly fallen meteorites~\citep{JewittGuilbert2012}. Other products of aqueous alteration, such as magnetite, have been detected on asteroids~\citep[e.g.][]{YangJewitt2010}, so it is possible that the absorptions observed in the spectrum of Themis are due to similar materials other than goethite.  If the H$_2$O ice is thermally decoupled from the regolith (i.e. it is clean and does not contain contaminating carbonaceous material), it will have a much lower equilibrium temperature due to the much higher albedo of water ice, meaning the sublimation rate will be much lower than we have assumed. This in turn would result in a much higher upper limit on the active fraction. In Fig.~\ref{sublimation}, we plot the derived active fraction of Themis as a function of the Bond albedo of the surface ice for several thermal models.  As we have an upper limit for the H$_2$O production rate and not a detection, each curve represents an upper bound to the region of permissable active fractions.  Even with an albedo of 0.99, our derived upper limit is still 10$^{-2}$ for the slow-rotator model. Therefore we favor an interpretation similar to that given by~\cite{JewittGuilbert2012}, where a small fraction of the surface is covered in ice, but this ice must be fairly clean.  More detailed evaluation of these possibilities is beyond the scope of this paper.\\

\indent The active fractions for both Themis and Ceres are much less than observed for comets, which are typically in the area of 1-10\%. This may be due either to an increased thickness for the insulating regolith on asteroid surfaces, or just a depletion of near-surface water ice relative to traditional comets.  It is also true that our sublimation models are probably oversimplifications, and the coverage of water ice on cometary nuclei visited by spacecraft is less than that predicted from sublimation models~\citep[e.g.][]{Sunshine2006}.  If we assume that the MBCs, such as 133P/Elst-Pizarro, have similar active fractions as inferred for Ceres and Themis, then MBCs would have H$_2$O production rates on the order of $<$ 10$^{24}$ mol s$^{-1}$ (assuming the slow-rotator model). These values are consistent with upper limits derived for MBCs so far~\citep{Licandro2011, deValBorro2012, ORourke2013} and would suggest that it will be very difficult to detect water sublimation directly around an MBC without a dedicated flyby/orbiting mission. However, estimates for the H$_2$O production rate for MBCs based on the observed dust coma are on the order of 10$^{25}$ mol s$^{-1}$, which should be detectable by the James Webb Space Telescope~\citep{Kelley2015}. If water vapor is detected around a Main Belt Comet at this production rate, then that suggests the active fractions of MBCs are larger than for larger asteroids like Themis or Ceres, but still less
than typical comets.\\

\indent Fig.~\ref{detlimit} shows the expected detection limit as a function of magnitude for APO (i.e. a 3-meter class telescope) and Keck (i.e. a 10-meter class telescope) for an object with Themis's approximate observing geometry.  This calculation is based on our results for Themis and is simplified in that it does not account for telescope specific differences (i.e. throughput of the instrument, etc.), only the difference in collecting area.  The decreasing limit with decreasing brightness is due to the decreasing strength of the continuum and its associated noise component.  Once the continuum is no longer detected, the detection limit flattens out to a terminal value.  At the faint end (i.e. MBCs) a 10-meter telescope can likely reach a detection limit of 10$^{26}$ mol s$^{-1}$.  We also overplot our derived upper limits for Ceres and Themis.  Themis trivially falls directly on the relation as this object was used to derive the plot.  Ceres is above the curve since the imperfect removal of a strong continuum introduces systematic scatter, whereas the relation was derived assuming purely Poisson statistics.  Therefore for bright objects (similar to Ceres and brighter) our estimates may be overly optimistic.\\ 

\section{Conclusions}

\indent We present upper limits on the water production from Main Belt asteroids 24 Themis and 1 Ceres. Our upper limit for Themis is the most constraining obtained to date, while our limit for Ceres is consistent with previous detections of water vapor. Our results suggest that less than a few $\times$ 10$^{-4}$ of the surface area for Themis and Ceres is actively sublimating. This implies active fractions at least 2-3 orders of magnitude less than comets. From analysis of our results
we favor a picture where water ice covers a small fraction of the surface area, but is relatively clean. The Dawn mission, currently in orbit around Ceres, will provide greater knowledge of the water ice distribution on Ceres. If the limits
on the active fraction derived here are applicable to MBCs as well, then this implies that outgassing rates should be $<$ 10$^{24}$ mol s$^{-1}$, making it difficult to directly detect without a dedicated spacecraft mission.  Our results show that searches for \ion{O}{1} emission around asteroids and other bodies suspected to be undergoing active sublimation can place meaningful constraints on the presence and properties of any surface ice present on these objects. 

%% Using an acknowledgements command is not in the Elsevier template,
%% but it can be used.
\ack
We thank two anonymous reviewers whose comments improved the quality of this manuscript.  We thank the APO observing staff for their invaluable help in conducting the observations.  We thank John Barentine, Jurek Krzesinski, Chris Churchill, Pey Lian Lim, Paul Strycker, and Doug Hoffman for developing and optimizing the ARCES IRAF reduction script used to reduce this data.  We would also like to acknowledge the JPL Horizons System, which was used to generate ephemerides for nonsidereal tracking of the asteroids during the ARCES observations, and the SIMBAD database, which was used for selection of reference stars.

\label{lastpage}

% Bibliographic references with the natbib package:
% Parenthetical: \citep{Bai92} produces (Bailyn 1992).
% Textual: \citet{Bai95} produces Bailyn et al. (1995).
% An affix and part of a reference:
%   \citep[e.g.][Ch. 2]{Bar76}
%   produces (e.g. Barnes et al. 1976, Ch. 2).-

\bibliography{../references.bib}

%% Use the plainnat style for ``Icarus'' mode to display DOI numbers
%% among other things.  However, revert to the Elsevier elsart-harv
%% mode for ``Elsevier'' mode.
\bibliographystyle{plainnat}

\clearpage

\begin{landscape}
\begin{table}
\caption{\textbf{Observation Log}
\label{observations}
\label{lasttable}
}
\begin{tabular}{llllllllll}
\hline
Object & Date (UT) & $r$ (AU) & $\Delta$ (AU) & $\dot{\Delta}$ (km s$^{-1}$) & True Anomaly (deg) & Days from Peri. & G2V & Fast Rot. & Flux Cal\\
\hline
Themis & 2/15/2013 & 2.89 & 2.56 & 21.7 & 303 & 258 & HD 25370 & HD 27660 & HR 1544\\
Themis & 2/23/2013 & 2.88 & 2.67 & 22.0 & 304 & 250 & HD 30455 & $\chi$ Tau & HR 1544\\
Ceres & 2/26/2013 & 2.62 & 2.22 & 21.2 & 311 & 201 & G105-14 & HR 2207 & HR 2207\\
\hline
\end{tabular}
\end{table}
\end{landscape}

\clearpage

\begin{table}
\caption{\textbf{Upper Limits for Fluxes and Production Rates}
\label{upperlimit}
\label{lasttable}
}
\begin{center}
\begin{tabular}{lllll}
\hline
Object & Flux (ergs s$^{-1}$ cm$^{-2}$) & Q$_{H_2O}$ (mol s$^{-1}$)\\
\hline
Themis & $<$ 6.2 $\times$ 10$^{-16}$ & $<$ 4.5 $\times$ 10$^{27}$\\
Ceres & $<$ 7.6 $\times$ 10$^{-15}$ & $<$ 4.6 $\times$ 10$^{28}$\\
\hline
\end{tabular}
\end{center}
\end{table}

\clearpage

\begin{landscape}
\begin{table}
\caption{\textbf{Upper Limits for Active Fractions}
\label{activefraction}
\label{lasttable}
}
\begin{center}
\begin{tabular}{ccccccc}
\hline
& \multicolumn{3}{c}{Themis} & \multicolumn{3}{c}{Ceres}\\
Model & Z (mol s$^{-1}$ cm$^{-2}$) & Active Area (km$^2$ & Active Fraction & Z (mol s$^{-1}$ cm$^{-2}$) & Active Area (km$^2$ & Active Fraction\\
\hline
Isothermal & 9.66 $\times$ 10$^{18}$ & $<$ 466 & $<$ 3.7 $\times$ 10$^{-3}$ & 3.58 $\times$ 10$^{19}$ & $<$ 1284 & $<$ 4.6 $\times$ 10$^{-4}$\\
Fast-Rotator & 2.18 $\times$ 10$^{19}$ & $<$ 206 & $<$ 1.6 $\times$ 10$^{-3}$ & 5.78 $\times$ 10$^{19}$ & $<$ 795 & $<$ 2.7 $\times$ 10$^{-4}$\\
Slow-Rotator & 1.86 $\times$ 10$^{20}$ & $<$ 24 & $<$ 1.9 $\times$ 10$^{-4}$ & 2.62 $\times$ 10$^{20}$ & $<$ 175 & $<$ 6.2 $\times$ 10$^{-5}$\\
Subsolar & 1.12 $\times$ 10$^{21}$ & $<$ 4 & $<$ 3.2 $\times$ 10$^{-5}$ & 1.46 $\times$ 10$^{21}$ & $<$ 31 & $<$ 1.1 $\times$ 10$^{-5}$\\
\hline
\end{tabular}
\end{center}
\end{table}
\end{landscape}

\begin{center}
Figure Captions
\end{center}

\begin{figure}[p!]
\begin{center}
\includegraphics[width=\linewidth]{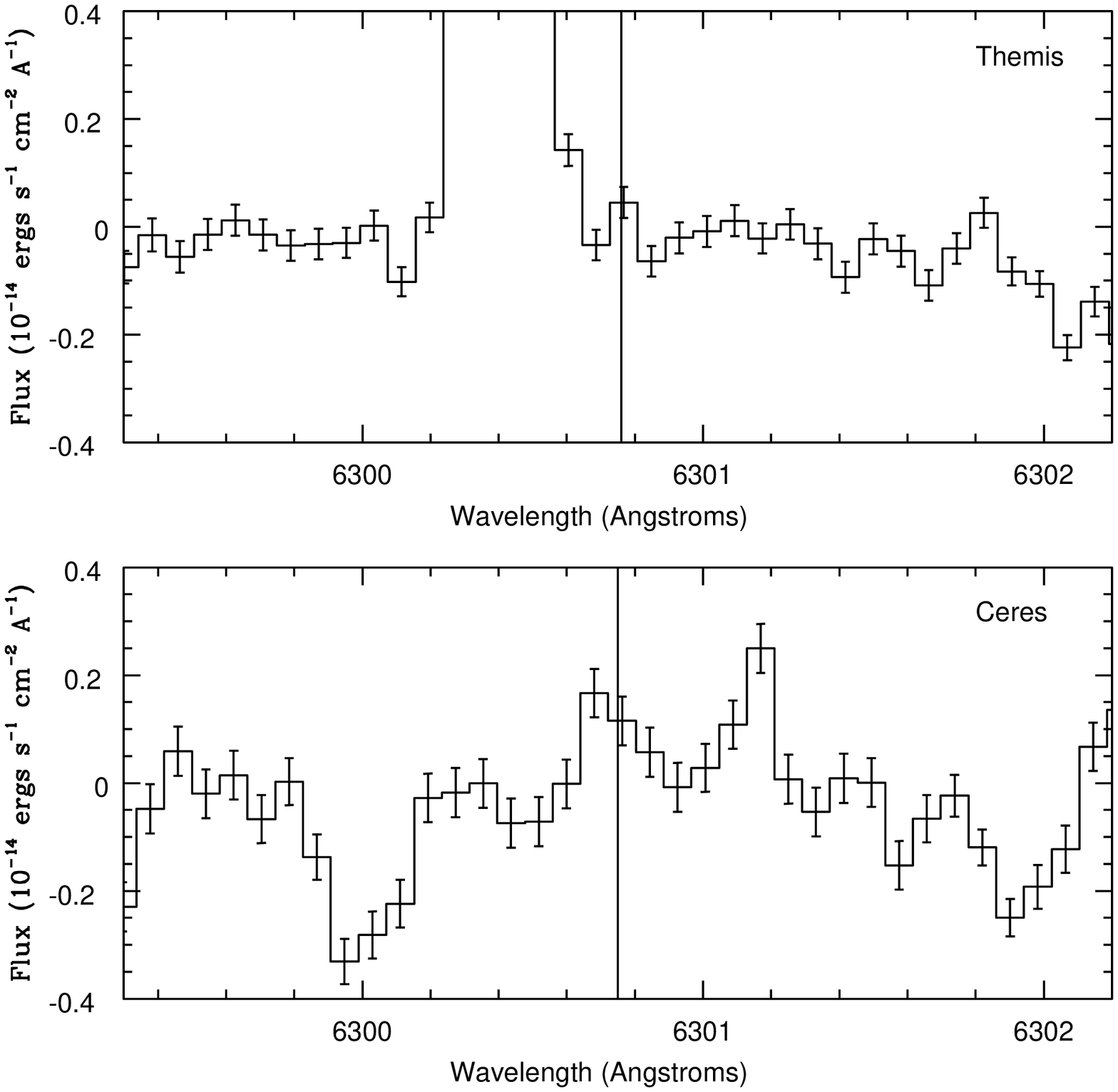}
\caption{
\label{spectra}
\label{lastfig}
}
\end{center}
\end{figure}

Fig~\ref{spectra}: Spectra of Themis (top) and Ceres (bottom) showing the spectral region of the [\ion{O}{1}]6300~\AA~line. The expected position of the asteroidal emission is indicated by the vertical line. The telluric feature is clearly observed for Themis and extends beyond the vertical scale of the plot, but there is no definitive evidence for [\ion{O}{1}]6300 emission from either Themis or Ceres.

\begin{figure}[p!]
\begin{center}
\includegraphics[width=\linewidth]{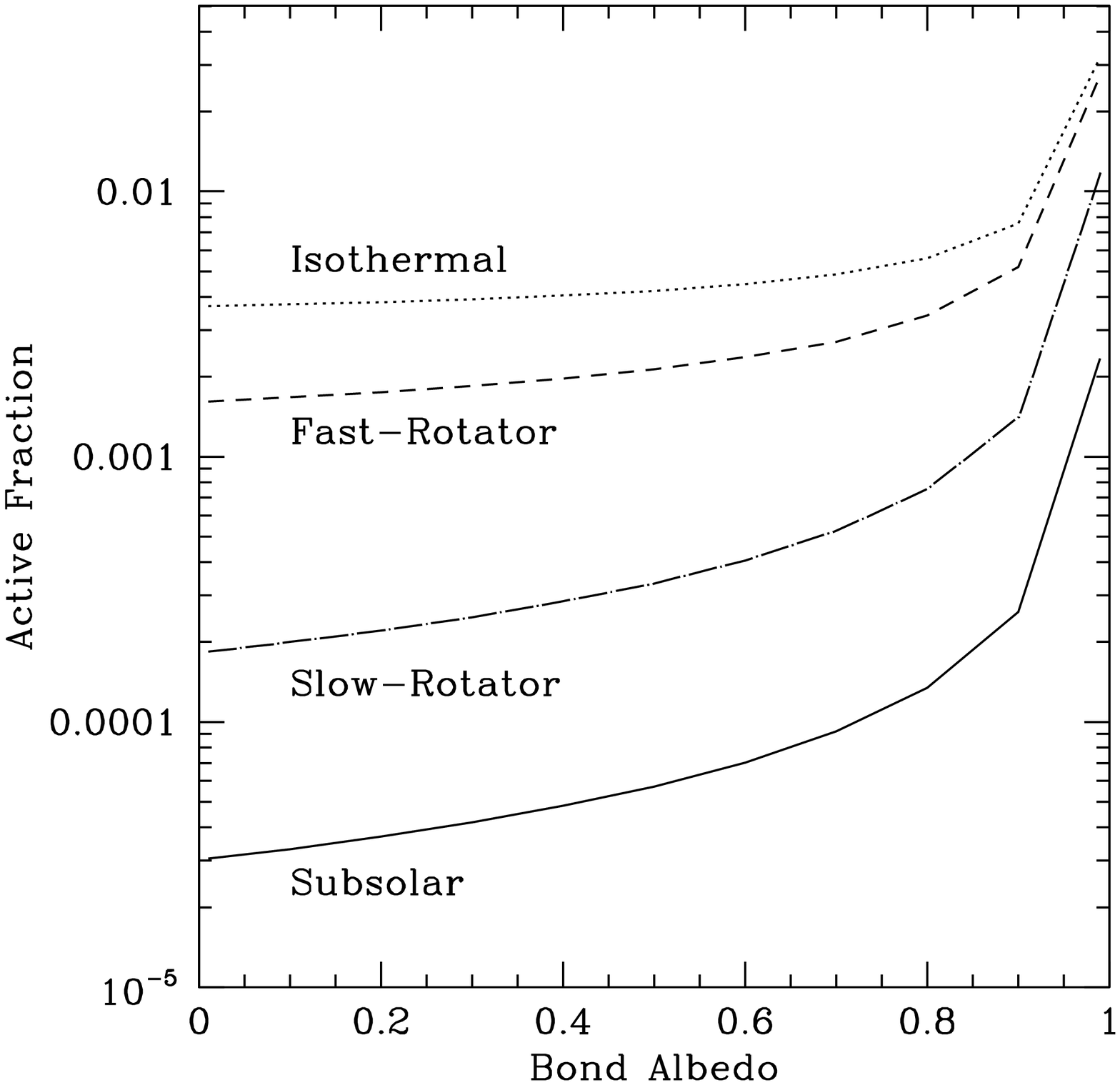}
\caption{
\label{sublimation}
\label{lastfig}
}
\end{center}
\end{figure}

Fig~\ref{sublimation}: Plot showing the active fraction of Themis as a function of Bond albedo of the surface water ice for different thermal models.  As our measurement only provides an upper limit on the active fraction, the curves represent upper bounds to the true active fraction.

\begin{figure}[p!]
\begin{center}
\includegraphics[width=\linewidth]{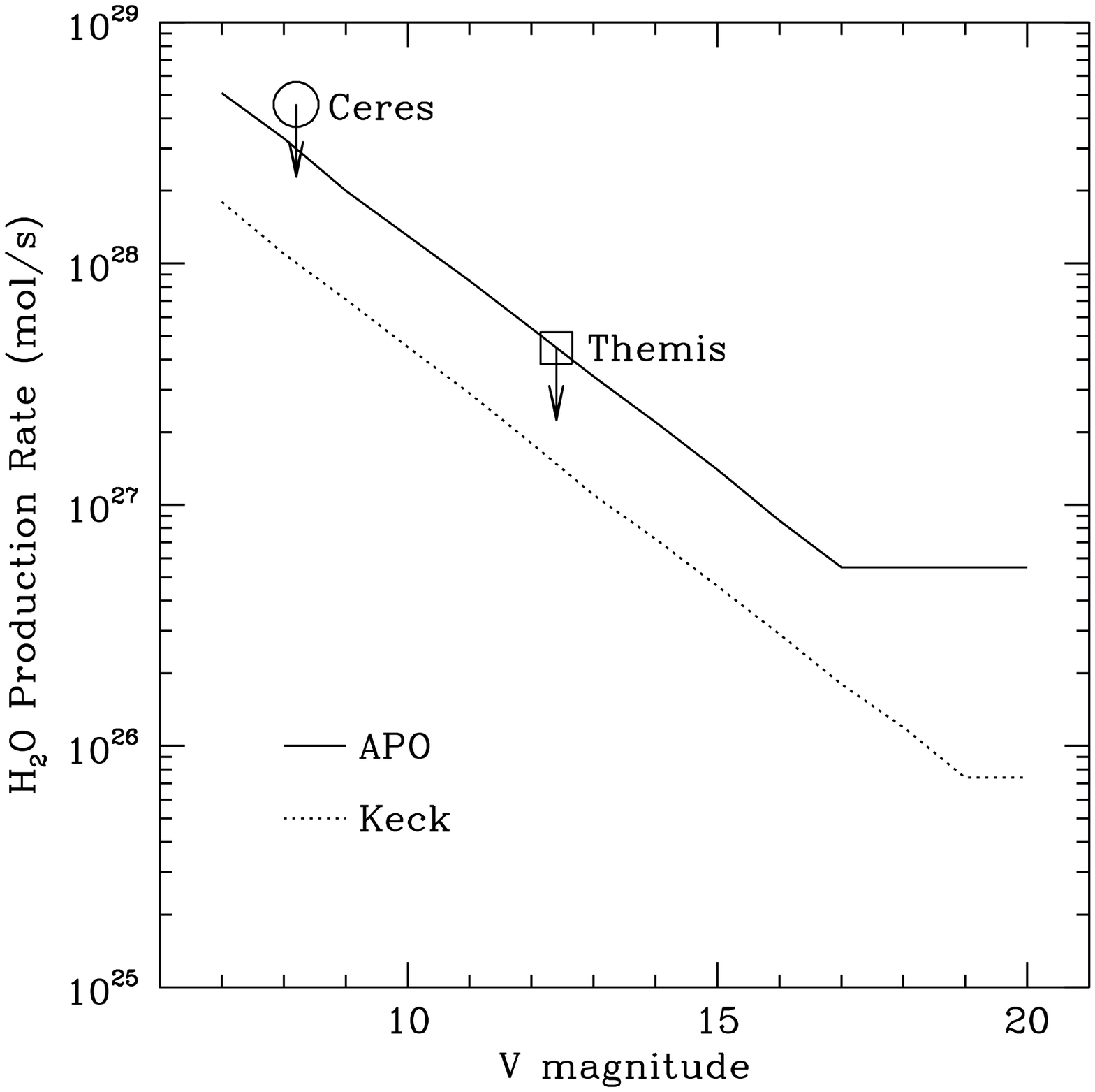}
\caption{
\label{detlimit}
\label{lastfig}
}
\end{center}
\end{figure}

Fig.~\ref{detlimit}: Approximate detection limits for H$_2$O using [\ion{O}{1}]6300 emission for a 3.5 meter (i.e. APO) and a 10 meter (i.e. Keck) telescope as a function of magnitude.  Limits become more sensitive for faint objects due to decreasing strength of the continuum.  The curves flatten out when continuum is no longer detected.  Our derived upper limits for Themis and Ceres are overplotted.  Themis falls directly on the curve as the Themis observations were used as the basis for the calculation of the detection limits.  Ceres is slightly above as our calculation does not account for the systematic scatter introduced when attempting to accurately subtract a strong continuum.  Therefore for very bright targets similar in brightness to Ceres our estimated detection limits may be overly optimistic. 

\clearpage

\clearpage

\clearpage

\end{document}